\documentclass[superscriptaddress,showpacs,aps,nofootinbib]{revtex4}
\usepackage{graphicx}
\usepackage{amsmath}

\newcommand {\eq}{\begin{equation}}
\newcommand {\ee}{\end{equation}}

\begin{document}

\title{Limits from Weak Gravity Conjecture on Dark Energy Models}
\date{}

\author{Xing Wu\footnote{wxxwwxxw@mail.bnu.edu.cn} and Zong-Hong Zhu\footnote{zhuzh@bnu.edu.cn}}

\address {Department of Astronomy, Beijing Normal University,
Beijing 100875, China}

\begin{abstract}
The weak gravity conjecture has been proposed as a criterion to
distinguish the landscape from the swampland in string theory. As an
application in cosmology of this conjecture, we use it to impose
theoretical constraint on parameters of two types of dark energy
models. Our analysis indicates that the Chaplygin-gas-type models
realized in quintessence field are in the swampland, whereas the $a$
power-low decay model of the variable cosmological constant can be
viable but the parameters are tightly constrained by the conjecture.
\end{abstract}

%\keywords{string theory, dark energy models, Chaplygin gas, variable cosmological constant }

\maketitle

\section{Introduction}
The observations of Type Ia supernovae \cite{SN} together with CMB
\cite{WMAP} and large scale structure \cite{SDSS} strongly indicate
that the expansion of the universe is accelerating. The fuel for the
cosmic acceleration is generally ascribed to an exotic component
with negative pressure, dubbed the dark energy, which accounts for
approximately $70\%$ in the content of the universe. The simplest
candidate for the dark energy is the cosmological constant or the
vacuum energy. Although most favored by all observations so far, it
suffers from the famous cosmological constant problem and the
coincidence problem (if one dose think of it as a problem)
\cite{cc}. That is, why the theoretical value of the vacuum energy
estimated from quantum field theory is enormously greater than that
observed? Why the energy density of the dark energy is of the same
orders of magnitude as that of matter? In fact, at cosmological
scales, the effect of gravity cannot be ignored when considering the
vacuum energy, therefore the nature of cosmological constant is
expected to be predicted authentically by some theory of quantum
gravity. On the other hand, besides the cosmological constant, there
are so many dark energy models which are highly degenerated in
fitting with observational data (see, for example, \cite{0603057}
for reviews). In other words, current observations fail to
definitely select one (or at least, a small number of) most probable
model(s). While waiting for future improved observations and new
scheme of detection, we also expect that theoretical development
would shed some light on this issue.

String theory is believed to be a consistent theory of quantum
gravity. However, one of its central problem is how to connect the
theory to experiments. For the superstring theory, which lives in
ten dimensions, the extra dimensions have to be compactified in
order to be relevant to the real world. A vast number of meta-stable
de Sitter vacua can be constructed through the scheme of flux
compactification on a Calabi-Yau manifold \cite{flux}. These string
vacua can be described by the low-energy effective field theories.
Recent researches \cite{swampland} indicate that a large amount of
these semi-classically consistent effective field theories are
essentially inconsistent at quantum level. These effective field
theories are in the so-called swampland, whereas the really
consistent ones are in the string landscape \cite{Susskind}.
Obviously it is of great significance to distinguish the landscape
from the swampland. Some criteria of consistent effective field
theory were proposed in \cite{swampland}. Recently the conjecture of
gravity as the weakest force proposed in \cite{WGC} further helps to
rule out those effective field theories in the swampland.  As
pointed in \cite{WGC}, when it comes to quantum gravity, gravity and
other gauge forces should not be considered separately. For a four
dimensional U(1) gauge field coupled to gravity with coupling $g$,
there naturally exists a new cut-off scale below the Planck scale in
asymptotic flat background: $\Lambda\sim gM_p$ where $M_p$ is the
Planck scale. Above this cut-off the effective field theory breaks
down and a more stringy approach is needed. This conjecture was
generalized to asymptotic dS/AdS background in \cite{dS}, where the
weak gravity conjecture together with the natural idea that the IR
cut-off should be smaller than the UV cut-off leads to an upper
bound for the cosmological constant $\rho_{\Lambda}\leq g^2M_p^4$.
Some evidence from string theory supporting this conjecture were
studied in \cite{string evidence}. As an application to cosmology,
this conjecture results in a new cut-off for the effective
$\lambda\phi^4$ theory for inflation, and it implies that the
chaotic inflation model is in the swampland \cite{chaotic}. It is
further conjectured in \cite{index} that the variation of the
inflaton should be less than $M_p$, which leads to the constraints
on the spectral index. Besides, the eternal chaotic inflation can
not be achieved when this conjecture is taken into account
\cite{eternal}. For the dark energy problem, by requiring that the
variation of the quintessence field be less than the Planck scale,
the equation of state (EoS) of quintessence can be tightly
constrained theoretically, and the result is consistent with
observations \cite{limit quint}.

In this paper, we are going to illustrate the theoretical limits on
the parameters of dark energy models. First of all, we introduce the
criterion inspired by the weak gravity conjecture. If we believe
that our universe is one of the vast landscape of vacua, then, as a
low-energy effective field theory to describe the vacuum energy, the
quintessence should not be in the swampland, namely, the variation
of the canonical scalar field should be less than the Planck mass
$M_p$. Following \cite{limit quint}, this means the expression \eq
{\Delta\phi(z_m)\over M_p}=\int_0^{z_m} {d\phi(z)\over
M_p}=\int_0^{z_m} \sqrt{3\left [1+w(z)\right ]{\rho\over
3M_p^2H^2}}{dz\over(1+z)} \label{criteria}<1,\ee where $w(z)$ and
$\rho$ respectively denote the EoS and the energy density of the
quintessence field.

To solve the dark energy problem,  one perspective is to invoke some
dynamic field, for which the EoS can be less than $-1/3$, as a
source to drive the acceleration: for quintessence, $-1<w<1$; for
phantom, $w<-1$; for tachyon field, $-1<w<0$. Beside these field
theory models, there are many other models built from a more
phenomenological perspective. A simple example is the quiessence
model which invokes a perfect fluid with negative pressure and a
constant EoS other than $-1$. What we are going to focus on here are
two classes of phenomenological models of dark energy: the
Chaplygin-gas-type models and the variable cosmological constant
(VCC) models. The phenomenological models can always be realized in
some more fundamental descriptions such as scalar field theories.
For example, the Chaplygin-gas-type models provide some particular
forms of EoS, as we will see below, which can always be used to
reconstruct the corresponding potentials for quintessence field or
other field theories.  In the following, we assume a flat FRW
universe consisting of dark energy and dust-like matter with the
Friedmann equation \eq 3M_p^2H^2=\rho_m+\rho_{DE}
\label{Friedmann}.\ee

The Chaplygin gas (CG) is a perfect fluid with the EoS \eq
p_{CG}=-{A\over\rho_{CG}},\label{EoS_CG}\ee where $A$ is a positive
constant. It was introduced by Chaplygin \cite{Chaplygin} in the
field of areodynamics. The Chaplygin gas model was first proposed in
\cite{CG} as an alternative to quintessence. Then it was used as a
unified description for dark matter and dark energy (UDME)
\cite{UDME}, where the UDME fluid evolves from a state of
pressureless dust in the past to the state like a cosmological
constant in the future. Such a EoS can be originated from tachyon
field described by the Born-Infeld action with constant potential
\cite{0603057,CG}, which can be related to a perturbed d-brane in
(d+2) dimensional spacetime.  A universe consisting of only the
Chaplygin gas can be realized by a quintessence field with the
potential \cite{CG}\eq V(\phi)={\sqrt{A}\over 2}\left
(\cosh\sqrt{3}\phi/M_p+{1\over{\cosh\sqrt{3}\phi/M_p}}\right ).\ee
When considering the existence of other component such as baryon or
radiation, the exact form of the potential will change, but in
principle, we can always reconstruct such a potential. The original
Chaplygin gas model is in fact incompatible with the observations.
As a generalization, the EoS (\ref{EoS_CG}) can be modified as
\cite{GCG} \eq p_{GCG}=-{A\over\rho_{GCG}^\alpha},\label{EoS_GCG}\ee
where $A$ and $\alpha$ are constants. In this case of the
Generalized Chaplygin Gas (GCG), the universe evolves from a dust
dominated phase, through a phase described by the EoS
$p=\alpha\rho$, to end up with a de Sitter phase. This EoS can be
derived from the generalized Born-Infeld theory \cite{GCG}. The
original EoS (\ref{EoS_CG}) can also be generalized to \eq
p_{VCG}=-{A(a)\over\rho_{VCG}},\label{EoS_VCG}\ee where $A(a)$ is
not a constant but a variable with respect to the scale factor $a$.
This variable Chaplygin gas (VCG) model is inspired by the
Born-Infled theories with potentials which are not constant
\cite{VCG}. It differs from the original CG in that it ends up in a
quiessence phase with the EoS $w_{VCG}=-1+6/n$. The GCG and the VCG
models seems compatible with some of current observations
\cite{Bento,Bertolami, Gong, ZHZ,VCG,VCG1,VCG2}. But there are also
some controversies over the compatibility of these models with some
other observational requirements \cite{counter}.

The VCC model is a more phenomenological description aimed at
solving the cosmological constant problem. In this scenario, the
cosmological constant decays with time $\Lambda=\Lambda(t)$ while
keeping the EoS $p_\Lambda(t)=-\rho_\Lambda(t)$. It is easy to see
that such forms of energy density and EoS generally require the
existence of interaction between dark energy and matter. There are
lots of works devoted to this issue with various decaying forms in
the literature (see \cite{sahni,overduin} for detailed lists of
these models). In this paper we consider one typical model with
\eq\Lambda=\beta a^{-m}\label{lambda},\ee where $m$ and $\beta$ are
positive constants. We will follow the method proposed in \cite{YZM}
to realize VCC in quintessence (see \cite{sahni,maia} for earlier
discussions on scalar description for VCC).

Before going to detailed analysis, two points should be noted.
First, in \cite{chaotic}, the weak gravity conjecture results in
$\phi\leq M_p$ for $V(\phi)=\lambda\phi^4$ and some other polynomial
potentials. In fact the minimum of such potential lies in $\phi_0=0$
, therefore the value of $\phi$ is essentially the variation with
respect to $\phi_0$. Moreover, the potential $V(\phi)$ can be
shifted without affecting physical results, and in general we do not
require the minimum position $\phi_0=0$. Thus the physically
meaningful quantity is indeed the variation of the field with
respect to some $\phi_0$, rather than its absolute magnitude.
Second, we choose the upper limit of the integral in
(\ref{criteria}) as $z_m=1089$, the recombination redshift, because
in general, the variation of $|\Delta\phi(z_m)|/M_p$ with respect to
$z_m$ is negligible for $z_m\gtrsim1000$, as we will see later.

The paper is organized as follows. The Chaplygin-gas-type models are
discussed in Section \ref{sec2}. The limits on the models of CG, GCG
and VCG are investigated respectively in the three subsections. In
Section \ref{sec3}, one typical model of VCC is studied. The final
section is devoted to the conclusion.

\section{Limits on the Chaplygin-gas-type models}\label{sec2}
\subsection{The Chaplygin gas model}
By energy conservation, we get the evolution of the energy density
of the CG \eq \rho_{CG}=\sqrt{A+{B\over a^6}}, \ee where $B$ is an
integration constant. Setting $a=a_0\equiv1$ leads to the initial
value $\rho_{CG0}=\sqrt{A+B}$. Defining $A_s=A/(A+B)$, the energy
density can be recast into \eq
\rho_{CG}=\rho_{CG0}\sqrt{A_s+(1-A_s)(1+z)^6} \label{rho_CG},\ee
where $1+z=1/a$ has been used. Then the EoS can be expressed by \eq
w_{CG}(z)=-{A_s\over A_s+(1-A_s)(1+z)^6},\label{w_CG}\ee  Note the
physical significance of $A_s$ is just $A_s=-w_{CG0}$, thus $A_s$
must be less than 1 in order that the Chaplygin gas can be realized
by a quintessence field and such that the criterion (\ref{criteria})
can be applied. By Eq.(\ref{rho_CG}) we have \eq {\rho_{CG}\over
3M_p^2H^2}={\rho_{CG}\over \rho_{CG}+\rho_m}={\Omega_{CG}\over
\Omega_{CG}+\Omega_m} \label{Omega_CG},\ee where
$\Omega_m=\Omega_{m0}(1+z)^3$ and $\Omega_{CG}={\rho_{CG}\over
3M_p^2H_0^2}$. Note that in the literature the parameter
$\Omega_{m0}$ often refers to the baryonic matter only. Even in some
literature (for example \cite{Bento,Bertolami}), the contribution
from baryon is just ignored. Here we make a loose assumption that
the matter component is not confined to baryonic matter, it may also
be a mixture of baryonic matter and the dark matter originated from
sources other than the Chaplygin-gas-type fluid.

\begin{figure}[htbp]
\begin{center}
\includegraphics[scale=.6]{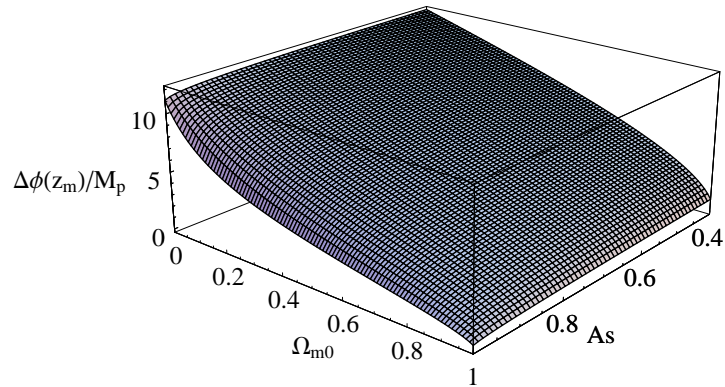} \caption[]{\small $\Delta\phi(1089)$ is far larger than $M_p$.
This implies that the original Chaplygin gas model can be ruled out
by the theoretical criterion.} \label{fig:CG}
\end{center}
\end{figure}

Applying the criterion (\ref{criteria}), we find that for all the
region of the parameter space, $\Delta\phi(1089)$ is far larger than
$M_p$, as we can see from Fig.\ref{fig:CG}. This implies that the
original Chaplygin gas model is inconsistent with the theoretical
requirement and therefore may not be a viable model. In fact,
constraints from observations also rule out the original CG model
\cite{Bento,Bertolami, Gong, ZHZ}.

\subsection{The generalized Chaplygin gas model}
\begin{figure}[htbp]
\begin{center}
\includegraphics[scale=.45]{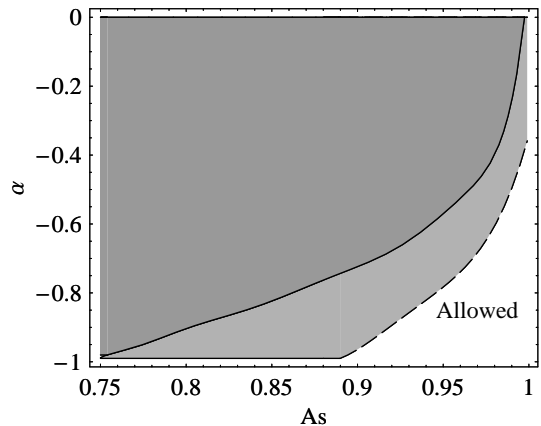} \caption[]{\small Limit on
the GCG parameter $A_s$ and $\alpha$ for $z_m=1089$. Below the
dashed line is the allowed region for $\Omega_{m0}=0.04$, and the
part under the solid line is the allowed region for
$\Omega_{m0}=0.27$. The two boundaries are obtained by setting
$|\Delta\phi(z_m)|=M_p$ in each cases. It can be seen that the
allowed range becomes enlarged for larger $\Omega_{m0}$.}
\label{fig:GCG}
\end{center}
\end{figure}

\begin{figure}[htbp]
\begin{center}
\includegraphics[scale=0.45]{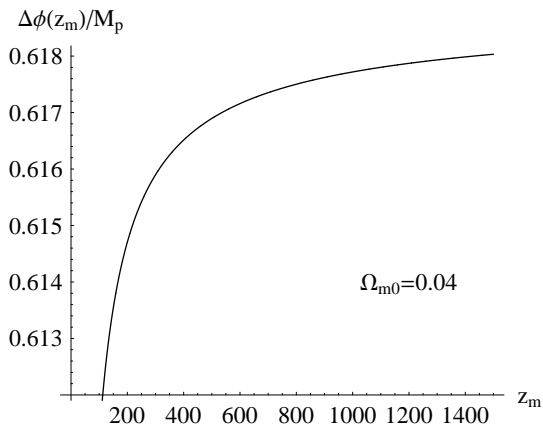} \caption[]{\small $|\Delta\phi(z_m)|/M_p$ as a function of
$z_m$. Its variation is negligible for $z_m\gtrsim 1000$.  }
\label{fig:deltaGCG}
\end{center}
\end{figure}

In this case, the basic equations are similar with those of the
original one. The energy density is \eq \rho_{GCG}=\rho_{GCG0}\left
[A_s+(1-A_s)(1+z)^{3(1+\alpha)}\right ]^{1/(1+\alpha)}
\label{rho_GCG}\ee and the EoS can be expressed by \eq
w_{GCG}(z)=-{A_s\over
A_s+(1-A_s)(1+z)^{3(1+\alpha)}}.\label{w_GCG}\ee Again,
$A_s=-w_{GCG0}$ should be less than unity. Note that when $\alpha=1$
we recover the original Chaplygin gas model. Defining the fractional
energy density the same way as in (\ref{Omega_CG}) and inserting
them into (\ref{criteria}), we get the limits on the GCG model. In
this model, there are in fact three parameters: $A_s$, $\alpha$ and
$\Omega_{m0}$. In Fig.\ref{fig:GCG} we present the theoretical
boundary on the $A_s$-$\alpha$ plane. We choose $\Omega_{m0}$ to be
$0.04$, $0.27$ respectively. The first case corresponds to
$\Omega_{m0}=\Omega_b$. That is, $\Omega_{m0}$ represents baryonic
matter only, and all the dark matter is described by the generalized
Chaplygin gas. And the value 0.04 is in accordance with observations
such as SDSS \cite{SDSS}, where $\Omega_bh^2=0.0222\pm0.0007$ and
$h=0.73\pm0.019$. The boundary of the allowed region is plotted with
the dashed line. In the second case, the GCG model is essentially a
model of dark energy without unifying dark matter, and the dark
matter contribution is included in $\Omega_{m0}$. The value 0.27 is
in accordance with WMAP \cite{WMAP}. The allowed part is bounded by
the solid line. Unlike the case of CG, GCG model can stand the
theoretical test within certain region of the parameter space as
shown in the figure. It is confirmed that CG, corresponding to
$\alpha=1$ on the plot, is far outside the allowed region. We also
illustrate in Fig.\ref{fig:deltaGCG} that the variation of
$|\Delta\phi(z_m)|/M_p$ is practically negligible for $z_m\gtrsim
1000$, in favor of that the choice of $z_m=1089$ is reasonable.

However, the allowed region is not consistent with observations. The
results of observational constraints are shown in Table
\ref{observation}. They all fall outside the theoretically allowed
region. This indicates that the GCG model realized in quintessence
is in the swampland. Since the figure shows that the greater
$\Omega_{m0}$ is, the larger the allowed region is, maybe GCG should
be considered as a model for only dark energy, instead of as UDME,
so that the allowed range may become large enough to be compatible
with observations. However, without the merit of unifying dark
matter and dark energy, this model is not as worthy as the simple
$\Lambda$CDM model, due to its introducing one more parameter.

\begin{table}
\begin{center}
\caption{\small Observational constraints} \label{observation}
\begin{tabular}{ccccc}
\hline
 $A_s$   & $\alpha$    & Reference& Observation & $\Omega_{m0}$     \\
\hline $[0.81,0.85]$       & $[0.2,0.6]$      &  \cite{Bento} &CMB  & $0$ \\
\hline
$0.70^{+0.16}_{-0.17}$   & $-0.09^{+0.54}_{-0.33}$ &\cite{ZHZ} & Galaxy cluster  & $\Omega_b$\\
 & & & + SNIa+FRIIb &  \\
\hline
$0.936$   & $3.75$      & \cite{Bertolami}       & SNIa         & $0$   \\
\hline $0.88^{+0.08}_{-0.03}$   & $1.57^{+0.1}_{-0.94}$  & \cite{Gong}    & SNIa     & $\Omega_b$\\
\hline

\end{tabular}
\end{center}
\end{table}

\subsection{The variable Chaplygin gas model}
\begin{figure}[htbp]
\begin{center}
\includegraphics[scale=0.6]{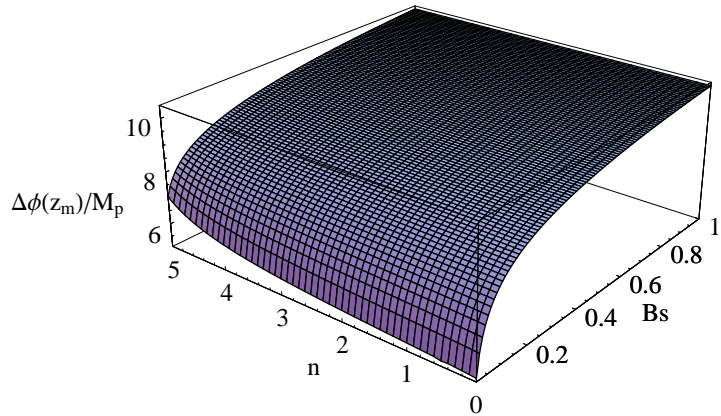} \caption[]{\small Limit on
the VCG parameter $B_s$ and $n$ for $z_m=1089$ and
$\Omega_{m0}=0.27$.  } \label{fig:VCG2}
\end{center}
\end{figure}

One difference of this model with CG and GCG is that it evolves into
a quiessence, rather than a cosmological constant as in CG and GCG.
By the energy conservation equation, the VCG density evolves as \eq
\rho_{VCG}=a^{-3}\left [6\int A(a)a^5da+B\right
]^{1/2},\label{rho_VCG_def}\ee where B is an integration constant.
Following \cite{VCG}, we assume the form $A(a)=A_0a^{-n}$, where
$A_0$ and $n$ are the parameters of the model. Inserting this form
into (\ref{rho_VCG_def}) leads to  \eq \rho_{VCG}=\sqrt{{6\over
6-n}{A_0\over a^n}+{B\over a^6}}.\label{rho_VCG}\ee From this
equation we can see that for $n<0$ the energy density increases with
time, exhibiting phantom behavior. So we impose $n\geq 0$. Note that
we can recover the original Chaplygin gas model for $n=0$. Setting
$a=a_0\equiv 1$ we get the initial value $\rho_{VCG0}=\sqrt{{6\over
6-n}A_0+B}$. Then defining $B_s=B/\rho_{VCG0}^2$ we can recast
(\ref{rho_VCG}) into a more useful form \eq
\rho_{VCG}=\rho_{VCG0}\sqrt{B_s(1+z)^6+(1-B_s)(1+z)^n},\ee from
which we obtain the range of $B_s$ is $0<B_s<1$. Then the EoS can be
expressed by \eq
w_{VCG}=-\frac{(6-n)/6(1-B_s)(1+z)^n}{B_s(1+z)^6+(1-B_s)(1+z)^n}.
\label{w_VCG}\ee By setting $z\rightarrow -1$ we can see that the
VCG ends up with a quiessence phase with $w_{VCG}=-1+n/6$. Following
the same way in above sections, we define the corresponding
fractional energy densities and insert them together with
(\ref{w_VCG}) into (\ref{criteria}). We find that when
$\Omega_{m0}=0.04$, the criterion is violated for all the given
range of the parameters. This indicates that the VCG realized in
quintessence as a unified description for dark energy and dark
matter is in the swampland. We expect the situation will be
ameliorated for $\Omega_{m0}=0.27$. But as is shown in
Fig.\ref{fig:VCG2}, the variation of the field is still greater than
$M_p$. Thus we conclude that the VCG can not be realized in a full
consistent quintessence field theory neither.

\section{Limits on the variable cosmological constant
model}\label{sec3}

As we mentioned before, VCC is a phenomenological description for
the cosmological constant decaying into matter fluid, and this
generally requires an interaction between matter and dark energy.
This scenario can be described by  the equations of continuity for
the two components as \eq
\dot{\rho}_\Lambda+3H(\rho_\Lambda+p_\Lambda)=-Q, \label{continue
lambda}\ee \eq \dot{\rho}_m+3H(\rho_m+p_m)=Q,\label{continue
matter}\ee with $Q$ phenomenologically introduced to denotes the
interaction. The interaction term effectively modified the equations
of state for matter and the cosmological constant. We assume the two
components only exchange pressure. For the cosmological constant,
its effective equation of state can be obtained by \eq w_{{\rm
eff}}=\frac{p_{\Lambda {\rm eff}}}{\rho_\Lambda}=-1+{m\over 3}, \ee
where $p_{\Lambda {\rm eff}}=p_\Lambda+Q/3H$. For matter although we
assume $w_m=0$, the relation $\Omega_m=\Omega_{m0}(1+z)^3$ does not
hold. It can be solved from (\ref{continue matter}). Then $H$ can be
obtained as \cite{YZM} \eq
H(z)=H_{0}[(\Omega _{m0}\frac{3}{3-m}-\frac{m}{3-m})(1+z)^{3}+\frac{3}{3-m}%
(1-\Omega _{m0})(1+z)^{m}]^{\frac{1}{2}}. \label{H}\ee By
(\ref{lambda}), we have $\rho_\Lambda=M_p^2\beta a^{-m}$. By setting
$z=0$ in the Friedmann equation we can express the coefficient
$\beta$ by $\beta=3H_0^2(1-\Omega_{m0})$. Inserting this expression
into (\ref{continue lambda}) leads to $Q=\beta mM_p^2(1+z)^mH$.

\begin{figure}[htbp]
\begin{center}
\includegraphics[scale=0.45]{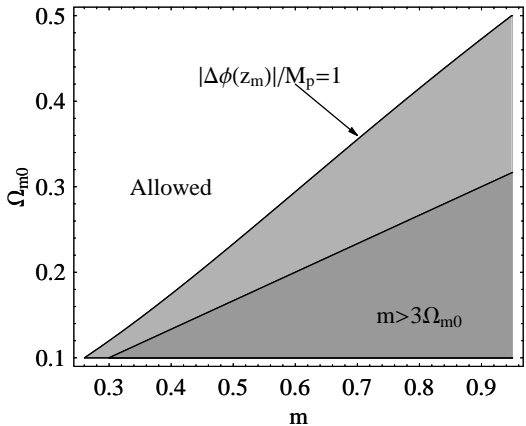} \caption[]{\small Limit on
the VCC parameter $m$ and $\Omega_{m0}$ . } \label{fig:VCC}
\end{center}
\end{figure}

Now we consider the scalar field description for this model. We use
$V_{{\rm eff}}$ to describe the effective potential containing the
contribution from the interaction $Q$. Then we have \eq {1\over
2}\dot{\phi}^2+V_{{\rm eff}}(\phi)=\rho_{\Lambda}, \label{rho Q}\ee
\eq {1\over 2}\dot{\phi}^2-V_{{\rm eff}}(\phi)=p_{\Lambda {\rm
eff}}.\label{p Q}\ee Combining these two equations leads to \eq
{d\phi\over
dz}=\mp\sqrt{\frac{mM_p^2(1-\Omega_{m0})(1+z)^{m-5}}{{3\Omega_{m0}-m\over{3-m}}+{3\over{3-m}}(1-\Omega_{m0})(1+z)^{m-3}}},\label{delta
VCC}\ee where $d/dt=-H(1+z)d/dz$ and (\ref{H}) have been used. Now
we can use (\ref{delta VCC}) to calculate the variation of the
quintessence field, and impose $|\Delta\phi|<M_p$ to obtain the
constraints on the parameters of this model. As shown in Fig.
\ref{fig:VCC}, the weak gravity conjecture sets an upper limit for
$m$ with given $\Omega_{m0}$. The condition $m<3\Omega_{m0}$ is
imposed by requiring that the expression in the square root of
(\ref{delta VCC}) should be positive. The model is constrained by SN
data in \cite{YZM}, where the best fit for $m$ is
$0.36^{+0.21}_{-0.23}$ at $1\sigma$, with $\Omega_{m0}=0.34$ given
as a priori. We can see clearly in the figure that this result is
within the allowed region.

\section{Conclusion}
In this paper, we use the criterion (\ref{criteria}) originated from
the weak gravity conjecture to investigate the feasibility of the
canonical scalar field description for two types of phenomenological
dark energy models. Although the models like the GCG and the VCG may
be compatible with observations (the CG is already ruled out by
observation), their theoretical foundation may be problematic. For
these models realized in quintessence, the CG and the VCG are in the
swampland. For the GCG, the criterion sets a very tight constraint
on the parameter space. However, this part is out of the best fit
range set by observations, indicating that the canonical scalar
field description of the GCG is incompatible with the theoretical
requirement (\ref{criteria}) either. Therefore, we reach the
conclusion that the Chaplygin-gas-type models can not be realized in
quintessence when the weak gravity conjecture is taken into account.
Whether these models can be described by the field theories like a
tachyon field or some generalized Born-Infeld theories is still an
open question worthy of further investigation. Besides, with a
particular form of decaying term (\ref{lambda}), we also illustrate
how to constrain another type of models, the VCC models. The
parameters of the model we use can be tightly constrained and the
method can be easily generalized to other VCC models as listed in
\cite{sahni,overduin}.

\section*{Acknowledgments}
We thank Yi-Fu Cai and Yin-Zhe Ma for helpful discussions. We are
also grateful to Neven Bili$\acute{\rm c}$, Gary Tupper and Raoul
Viollier for pointing out some of our misunderstandings. This work
was supported by the National Natural Science Foundation of China,
under Grant No.10533010, 973 Program No.2007CB815401 and Program for
New Century Excellent Talents in University (NCET) of China.

\end{document}